\documentclass[aps, rpl, preprint, superscriptaddress, showpacs]{revtex4}
\usepackage{epsfig,amsmath}

\begin{document}

\title
{Quantum nonlocality of four-qubit entangled states}
\author{Chunfeng Wu}
\affiliation{Department of Physics, National University of Singapore, 2 Science Drive 3, Singapore 117542}
\affiliation{Institute of Theoretical Physics, Northeast Normal University, Changchun 130024, P. R. China}
\author{Ye Yeo}
\affiliation{Department of Physics, National University of Singapore, 2 Science Drive 3, Singapore 117542}
\author{L. C. Kwek}
\affiliation{Department of Physics, National University of Singapore, 2 Science Drive 3, Singapore 117542}
\affiliation{Nanyang Technological University, National Institute of Education, 1 Nanyang Walk, Singapore 637616}
\author{C. H. Oh}
\email{phyohch@nus.edu.sg}
\affiliation{Department of Physics, National University of Singapore, 2 Science Drive 3, Singapore 117542}

\begin{abstract}
Quantum nonlocality of several four-qubit states is investigated by constructing a new Bell inequality.  These include the Greenberger-Zeilinger-Horne (GHZ) state, W state, cluster state, and the state $|\chi\rangle$ that has been recently proposed in [PRL, {\bf 96}, 060502 (2006)].  The Bell inequality is optimally violated by $|\chi\rangle$ but not violated by the GHZ state.  The cluster state also violates the Bell inequality though not optimally.  The state $|\chi\rangle$ can thus be discriminated from the cluster state by using the inequality.  Different aspects of four-partite entanglement are also studied by considering the usefulness of a family of four-qubit mixed states as resources for two-qubit teleportation.  Our results generalize those in [PRL, {\bf 72}, 797 (1994)].
\end{abstract}

\maketitle

\section{Introduction}\label{1}
Since Schrodinger's seminal paper in 1935 \cite{Schrodinger}, entanglement is recognized as being at the heart of quantum mechanics.  It engenders correlations between quantum systems much stronger than any classical correlation could be \cite{Bell, CHSH}.  Recently, entanglement has also been recognized as an essential physical resource in quantum information processing \cite{Nielsen}.  The power of entanglement in quantum communication can most convincingly be demonstrated by teleportation.  In their 1993 paper \cite{Bennett}, Bennett {\em et al.} have shown that entanglement shared between Alice and Bob can be used to teleport an unknown quantum state.  Slightly more than four years later, Bouwmeester {\em et al.} \cite{Bouwmeester} reported the first experimental demonstration of quantum teleportation.  In their experiment, they produced the necessary pairs of entangled photons by the process of parametric down conversion.  An important issue, which determines the success of their experiment, is thus whether or not the produced state is entangled.  In many of the experiments in quantum information science, entanglement witnesses are used for entanglement verification.  A violation of a Bell inequality can formally be expressed as a witness for entanglement \cite{Terhal}, and hence a good candidate for that purpose.

In addition to practical importance, quantum teleportation provides a useful theoretical framework to study entanglement.  For instance, Popescu \cite{Popescu} explored the different aspects of entanglement by analyzing the ``usefulness'' of Werner (channel) states \cite{Werner}
\begin{equation}
\rho_{\rm W} = q|\Psi_{\rm Bell}\rangle\langle\Psi_{\rm Bell}| + \frac{1 - q}{4}I_4,
\label{Werner}
\end{equation}
as resources for single-qubit teleportation.  Here, $0 \leq q \leq 1$, $|\Psi_{\rm Bell}\rangle \equiv (|00\rangle + |11\rangle)/\sqrt{2}$, and $I_4$ is the four-dimensional identity.  $\rho_{\rm W}$ can be useful resources for the standard teleportation protocol ${\cal S}_0$ of Bennett {\it et al.} \cite{Bennett} when $q > q_{\rm crit}({\cal S}_0) = 1/3$ \cite{Horodecki1}.  Clearly some of these states do not violate the Clauser-Horne-Shimony-Holt (CHSH) inequality \cite{CHSH}, since to do so demands $q > q_{\rm crit}({\rm Bell}) = 1/\sqrt{2}$ \cite{Horodecki2}.  The {\em critical visibility} $q_{\rm crit}({\rm Bell})$ measures the strength of Bell-inequality violation \cite{Werner}.  It is the minimum amount $q$ of a given entangled state $|\Psi\rangle$ that one has to add to white noise, so that the resulting state violates local realism.  The quantity $q_{\rm crit}({\rm Bell})$ is thus the threshold visibility above which the state cannot be described by local realism.

Recently, Yeo and Chua \cite{YeoI} presented an explicit protocol ${\cal E}_0$ for faithfully teleporting an arbitrary two-qubit state via a genuine four-qubit entangled state,
\begin{equation}
|\chi\rangle = \frac{1}{2\sqrt{2}}
(|0000\rangle-|0011\rangle-|0101\rangle+|0110\rangle +|1001\rangle+|1010\rangle+|1100\rangle+|1111\rangle).
\label{chi}
\end{equation}
This ``maximally" entangled state belongs to the following family of states
\begin{equation}
|\Upsilon^{00}(\theta_{12}, \phi_{12})\rangle = \frac{1}{\sqrt{2}}
(|\zeta^0(\theta_{12}, \phi_{12})\rangle + |\zeta^1(\theta_{12}, \phi_{12})\rangle),
\end{equation}
with
$$
|\zeta^0\rangle \equiv \frac{1}{\sqrt{2}}
(\cos\theta_{12}|0000\rangle - \sin\theta_{12}|0011\rangle - \sin\phi_{12}|0101\rangle + \cos\phi_{12}|0110\rangle)
$$
and
$$
|\zeta^1\rangle \equiv \frac{1}{\sqrt{2}}
(\cos\phi_{12}|1001\rangle + \sin\phi_{12}|1010\rangle + \sin\theta_{12}|1100\rangle + \cos\theta_{12}|1111\rangle).
$$
Here, $\theta_{12} \equiv \theta_1 - \theta_2$, $\phi_{12} \equiv \phi_1 - \phi_2$, and $0 < \theta_1, \theta_2, \phi_1, \phi_2 < \pi/2$.  When $\theta_{12} = \phi_{12} = \pi/4$, Eq.(3) reduces to Eq.(2).  In Ref.\cite{YeoII}, one of us considered teleportation with a mixed state of four qubits and defined the {\em generalized singlet fraction}.

Multipartite entanglement, still under intensive research, is not a direct extension of the bipartite case.  For instance, four qubits can be entangled in at least nine different ways \cite{Verstraete, Osterloh}.  It is thus insufficient to just say if a given state is entangled, it is also necessary for one to discriminate one entangled state from another.  Two types of Bell inequalities have been proposed for four qubits.  The well-known one is the 4-qubit Mermin-Ardehali-Belinskii-Klyshko (MABK) inequality \cite{Mermin90, Ardehali, BK}.  It is optimally violated by the 4-qubit GHZ state \cite{Greenberger}.  Recently, Scarani, Acin, Schenck, and Aspelmeyer \cite{cluster} proposed another Bell inequality for four qubits.  Here, we call it the SASA inequality.  It is not violated by the GHZ state, but optimally violated by the cluster state \cite{Briegel}.  Therefore, the SASA inequality allows one to discriminate between GHZ and cluster states.

In anticipation of a future experimental implementation of the above teleportation protocol we derive, in this paper, a new four-qubit Bell inequality that is optimally violated by the state $|\chi\rangle$.  This, together with results obtained in Ref.\cite{YeoII}, also enables us to carry out a study of the different aspects of multipartite entanglement, similar to that performed by Popescu \cite{Popescu}.  Our results show that nonlocality is more fragile to teleport than entanglement, and also generalize Popescu's results.  That is, there are ``local'' four-qubit states, which are nevertheless useful resources for ${\cal E}_0$.

Our paper is organized as follows.  In Section \ref{2}, we study the quantum nonlocality of $|\chi\rangle$ using the four-qubit MABK and SASA inequalities.  We show that $|\chi\rangle$ violates both inequalities.  However, the degrees of violation are $4\sqrt{2}$ and $2\sqrt{2}$ respectively, which are not optimal.  In Section \ref{3}, we first describe the formulation of our new Bell inequality.  Next, we show that it is optimally violated by $|\chi\rangle$.  This is followed by an analysis of the nonlocality of four-qubit GHZ, W \cite{Zeilinger} and cluster states using our Bell inequality.  It is found that the new Bell inequality is a good candidate for testing quantum nonlocality of the state $|\chi\rangle$ experimentally.  In Section \ref{4}, we explore different aspects of four-partite entanglement by analyzing the ``usefulness'' of the state $\xi$ [Eq.(\ref{Xi})] as a resource for two-qubit teleportation.  Lastly, quantum nonlocality of the state $|\Upsilon^{00}(\theta_{12}, \phi_{12})\rangle$ is also investigated in Section \ref{5}.  We summarize our results in the last Section \ref{6}.

\section{Previous Bell inequalities}\label{2}
The first Bell inequality for four qubits was derived by Mermin \cite{Mermin90}, Ardehali \cite{Ardehali}, Belinskii and Klyshko \cite{BK}.  Consider four observers: Alice $(A)$, Bob $(B)$, Charlie $(C)$ and Diana $(D)$, each having one of the qubits.  The formulation of the MABK inequality is based on the assumption that every observer is allowed to choose between two dichotomic observables.  Denote the outcome of observer $X$'s measurement by $X_i$ ($X = A, B, C, D$), with $i = 1, 2$.  Under the assumption of local realism, each outcome can either take value $+1$ or $-1$.  In a specific run of the experiment, the correlations between the measurement outcomes of all four observers can be represented by the product $A_iB_jC_kD_l$, where $i, j, k, l = 1, 2$.  In a local realistic theory, the correlation function of the measurements performed by all four observers is the average of $A_iB_jC_kD_l$ over many runs of the experiment:
\begin{equation}
Q(A_iB_jC_kD_l) = \langle A_iB_jC_kD_l\rangle.
\end{equation}
The MABK inequality reads \cite{Mermin90,Ardehali,BK}
\begin{eqnarray}
&   &  Q_{1111} - Q_{1112} - Q_{1121} - Q_{1211} - Q_{2111}            \nonumber  \\
& - &  Q_{1122} - Q_{1212} - Q_{2112} - Q_{1221} - Q_{2121} - Q_{2211} \nonumber  \\
& + &  Q_{2222} + Q_{2221} + Q_{2212} + Q_{2122} + Q_{1222} \leq 4,
\end{eqnarray}
where $Q_{ijkl}$ is short for $Q(A_iB_jC_kD_l)$.

In a quantum mechanical description, each observer $X$ measures the spin of each qubit by projecting it either along $\hat{n}^X_1$ or $\hat{n}^X_2$.  Every observer can independently choose between two arbitrary directions.  For the four-qubit state $|\chi\rangle$, the correlation functions are thus given by
\begin{equation}
Q(A_iB_jC_kD_l) = \langle\chi|\hat{n}^A_i\cdot\vec{\sigma} \otimes \hat{n}^B_j\cdot\vec{\sigma} \otimes
\hat{n}^C_k\cdot\vec{\sigma} \otimes \hat{n}^D_l\cdot\vec{\sigma}|\chi\rangle.
\label{cfbar}
\end{equation}
That is, the correlation functions are the expectation values of the joint two-outcome measurements on $|\chi\rangle$.  Here, $\vec{\sigma} = \sigma_x\hat{x} + \sigma_y\hat{y} + \sigma_z\hat{z}$; and $\sigma_x$, $\sigma_y$, and $\sigma_z$ are the Pauli matrices.  Under the experimental settings: $\hat{n}^A_1 = \hat{x}$, $\hat{n}^A_2 = \hat{z}$;  $\hat{n}^B_1 = \hat{y}$, $\hat{n}^B_2 = \hat{z}$; $\hat{n}^C_1 =\hat{y}$, $\hat{n}^C_2 = \hat{z}$; and $\hat{n}^D_1 = (\hat{z}-\hat{x})/\sqrt{2}$, $\hat{n}^D_2 = (\hat{x}+\hat{z})/\sqrt{2}$; we obtain the quantum prediction for the left hand side of the MABK inequality to be $4\sqrt{2}$.  Hence, $|\chi\rangle$ violates the MABK inequality.  We note that cluster states yield the same violation of the MABK inequality \cite{cluster}.

Next, we analyze the nonlocal property of $|\chi\rangle$ using the four-qubit SASA inequality proposed in Ref.\cite{cluster}, which can be cast in the following simple form,
\begin{equation}
Q(A_2B_1C_1D_1) + Q(A_1C_1D_2) + Q(A_1C_2D_1) - Q(A_2B_1C_2D_2) \leq 2,
\label{clusterbi}
\end{equation}
where $Q(A_iC_kD_l)$ is the correlation function of the measurements when Bob does not perform any measurement on his qubit.  It is noteworthy that there is only one local setting for one of the four observers (Bob) in the formulation of the SASA inequality.  This is in contrast to most Bell inequalities, which are constructed based on the assumption of two local settings for each observer (see, for instance, Refs.\cite{Mermin90,Ardehali,BK,cluster,ZB,werner,3qubit}).  Quantum mechanically,
\begin{equation}
Q(A_iC_kD_l) = \langle\chi|
\hat{n}^A_i\cdot\vec{\sigma} \otimes{\bf 1}^B \otimes \hat{n}^C_k\cdot\vec{\sigma} \otimes \hat{n}^D_l\cdot\vec{\sigma}|\chi\rangle.
\end{equation}
By appropriately choosing the following experimental settings: $\hat{n}^A_1 = \hat{x}$, $\hat{n}^A_2 = \hat{z}$; $\hat{n}^B_1 = \hat{z}$; $\hat{n}^C_1 =(\hat{x} + \hat{z})/\sqrt{2}$, $\hat{n}^C_2 = (\hat{x} - \hat{z})/\sqrt{2}$; and $\hat{n}^D_1 = \hat{z}$, $\hat{n}^D_2 = \hat{z}$; we determine the quantum prediction for the left hand side of the inequality (\ref{clusterbi}) to be $1/\sqrt{2} + 1/\sqrt{2} + 1/\sqrt{2} - (-1/\sqrt{2}) = 2\sqrt{2}$, which is greater than $2$.

The conflict between local realism and quantum mechanics is therefore obvious, but $|\chi\rangle$ does not optimally violate both the MABK and SASA inequalities.  In the next section, we will formulate a new four-qubit Bell inequality that is maximally violated by $|\chi\rangle$.

\section{The optimal Bell inequality}\label{3}
In contrast to the SASA inequality, now we suppose that Alice (instead of Bob) is only allowed to choose a single dichotomic observable parameterized by $\hat{n}^A_1$.  The other observers continue to choose independently between two arbitrary dichotomic observables parameterized by $\hat{n}^X_1$ and $\hat{n}^X_2$, with $X = B, C, D$.  Consequently, we need only to consider the correlation functions
\begin{equation}
Q(A_1, B_j, C_k, D_l) = \langle A_1B_jC_kD_l\rangle,
\label{cf1}
\end{equation}
and
\begin{equation}
Q(B_j, C_k, D_l) = \langle B_jC_kD_l\rangle.
\label{cf2}
\end{equation}
The following identity holds for the predetermined results:
\begin{equation}
A_1B_1C_1D_1 + B_1C_2D_2 + B_2C_1D_2 - A_1B_2C_2D_1 = \pm 2.
\label{quantity}
\end{equation}
Equation (\ref{quantity}) can be proved by direct enumeration of all the possible values that $X_i$ can take.  We rewrite the left hand side as follows: $A_1D_1(B_1C_1 - B_2C_2) + (B_1C_2 + B_2C_1)D_2$.  Since $X_i = \pm 1$, we know that $A_1D_1 = \pm 1$ and $D_2 = \pm 1$.  For the other two terms $B_1C_1 - B_2C_2$ and $B_1C_2 + B_2C_1$, it can be calculated that
\begin{eqnarray}
B_1C_1 - B_2C_2 = 0 & {\rm and} & B_2C_1 + B_1C_2 = \pm 2 \nonumber
\end{eqnarray}
or
\begin{eqnarray}
B_1C_1 - B_2C_2 = \pm 2 & {\rm and} &B_2C_1 + B_1C_2 = 0. \nonumber
\end{eqnarray}
So, $A_1B_1C_1D_1 + B_1C_2D_2 + B_2C_1D_2 - A_1B_2C_2D_1$ is either $+2$ or $-2$.  After averaging over many runs of the experiment, one can use the correlation functions defined in Eqs.(\ref{cf1}) and (\ref{cf2}) to express the left hand side of the identity, and obtain the following Bell inequality
\begin{equation}
Q(A_1B_1C_1D_1) + Q(B_1C_2D_2) + Q(B_2C_1D_2) - Q(A_1B_2C_2D_1) \leq 2.
\label{bi1}
\end{equation}
We note that through cyclic permutation of the four observers, $A \rightarrow B \rightarrow C \rightarrow D \rightarrow A$, we can derive from (\ref{bi1}) the following inequality
\begin{equation}
Q(A_1B_1C_1D_1) + Q(A_2C_1D_2) + Q(A_2C_2D_1) - Q(A_1B_1C_2D_2) \leq 2,
\label{bi2}
\end{equation}
which is equivalent to the SASA inequality (\ref{clusterbi}).  However, we must emphasize that since the entangled state $|\chi\rangle$ and the cluster state are not invariant under all possible permutations of the qubits, interchanging Alice and Bob does give rise to observable difference.  In fact, as we shall see below, $|\chi\rangle$ and the cluster state do yield different violations of the two inequalities.

Now, we will use the inequality (\ref{bi1}) to test the quantum nonlocality of $|\chi\rangle$.  Quantum mechanically, we have
\begin{eqnarray}
Q(A_1B_jC_kD_l) & = & \langle\chi|
\vec{n}^A_1\cdot\vec{\sigma}\otimes\vec{n}^B_j\cdot\vec{\sigma}\otimes\vec{n}^C_k\cdot\vec{\sigma}\otimes\vec{n}^D_l\cdot\vec{\sigma}
|\chi\rangle, \nonumber \\
Q(B_jC_kD_l) & = & \langle\chi|
{\bf 1}^A\otimes\vec{n}^B_j\cdot\vec{\sigma}\otimes\vec{n}^C_k\cdot\vec{\sigma}\otimes\vec{n}^D_l\cdot\vec{\sigma}|\chi\rangle.
\end{eqnarray}
We observe that $|\chi\rangle$ satisfies
\begin{eqnarray}
\sigma^A_x\sigma^B_z\sigma^C_z\sigma^D_x|\chi\rangle & = & |\chi\rangle, \nonumber \\
\sigma^A_x\sigma^B_x{\bf 1}^C\sigma^D_z|\chi\rangle  & = & |\chi\rangle, \nonumber \\
{\bf 1}^A\sigma^B_x\sigma^C_x{\bf 1}^D|\chi\rangle   & = & |\chi\rangle.
\end{eqnarray}
By multiplying the above three equations using the algebra of Pauli matrices, we obtain
\begin{eqnarray}
{\bf 1}^A\sigma^B_y\sigma^C_z\sigma^D_y|\chi\rangle  & = &  |\chi\rangle, \nonumber \\
\sigma^A_x\sigma^B_y\sigma^C_y\sigma^D_x|\chi\rangle & = & -|\chi\rangle, \nonumber \\
{\bf 1}^A\sigma^B_z\sigma^C_y\sigma^D_y|\chi\rangle  & = &  |\chi\rangle.
\end{eqnarray}
From the above six equations, we choose four terms and combine them as follows,
\begin{equation}
(\sigma^A_x\sigma^B_z\sigma^C_z\sigma^D_x + {\bf 1}^A\sigma^B_y\sigma^C_z\sigma^D_y + {\bf 1}^A\sigma^B_z\sigma^C_y\sigma^D_y
- \sigma^A_x\sigma^B_y\sigma^C_y\sigma^D_x)|\chi\rangle = 4|\chi\rangle.
\end{equation}
Therefore, with the following suitably chosen measurement settings:
\begin{eqnarray}
& & \vec{n}^A_1 = \vec{x}; \nonumber \\
& & \vec{n}^B_1 = \vec{z},\;\;\;\;\;\; \vec{n}^B_2 = \vec{y}; \nonumber \\
& & \vec{n}^C_1 = \vec{z},\;\;\;\;\;\; \vec{n}^C_2 = \vec{y};\nonumber \\
& & \vec{n}^D_1 = \vec{x},\;\;\;\;\;\; \vec{n}^D_2 = \vec{y};
\label{setting}
\end{eqnarray}
the left hand side of the inequality (\ref{bi1}) is $4$.  We construct a Bell quantity from the inequality (\ref{bi1})
\begin{equation}
{\cal B}^{\chi}_{(\ref{bi1})} = 
\langle\chi|(\sigma^A_x\sigma^B_z\sigma^C_z\sigma^D_x + {\bf 1}^A\sigma^B_y\sigma^C_z\sigma^D_y + {\bf 1}^A\sigma^B_z\sigma^C_y\sigma^D_y - \sigma^A_x\sigma^B_y\sigma^C_y\sigma^D_x)|\chi\rangle = 4.
\end{equation}
The correlation functions $Q(A_1B_jC_kD_l)$ and $Q(B_jC_kD_l)$ can take value either $+1$ or $-1$ under both local realistic theory and quantum mechanical theory.  Thus, the maximum value of the combination $|Q(A_1B_1C_1D_1) + Q(B_2C_1D_2) + Q(B_1C_2D_2) - Q(A_1B_2C_2D_1)|$ is $4$.  The above quantum prediction value of $4$ is thus the optimal violation of the inequality (\ref{bi1}).  There is no other state that can give a higher violation.

We close this section with a few remarks.  First, the four-qubit GHZ state $|\Psi_{\rm GHZ}\rangle = (|0000\rangle + |1111\rangle)/\sqrt{2}$ does not violate the inequality (\ref{bi1}).  The correlation functions of spin-component measurements on the GHZ state are calculated as follows,
\begin{equation}
Q^{\rm GHZ}(A_1B_jC_kD_l) = \langle\Psi_{\rm GHZ}|\vec{n}^A_1\cdot\vec{\sigma} \otimes \vec{n}^B_j\cdot\vec{\sigma} 
\otimes \vec{n}^C_k\cdot\vec{\sigma} \otimes \vec{n}^D_l\cdot\vec{\sigma}|\Psi_{\rm GHZ}\rangle
\end{equation}
and
\begin{equation}
Q^{\rm GHZ}(B_jC_kD_l) =  \langle\Psi_{\rm GHZ}|
{\bf 1}^A \otimes \vec{n}^B_j\cdot\vec{\sigma} \otimes \vec{n}^C_k\cdot\vec{\sigma} \otimes \vec{n}^D_l\cdot\vec{\sigma}
|\Psi_{\rm GHZ}\rangle = 0.
\end{equation}
Since $Q^{\rm GHZ}(A_1B_jC_kD_l)$ can only be $\pm 1$, it is clear that the Bell quantity ${\cal B}^{\rm GHZ}_{(\ref{bi1})} = Q^{\rm GHZ}(A_1B_1C_1D_1) - Q^{\rm GHZ}(A_1B_2C_2D_1)$ is never greater than $2$, which means that the inequality (\ref{bi1}) is not violated by the GHZ state.  Next, for the four-qubit W state $|\Psi_{\rm W}\rangle = (|1000\rangle+|0100\rangle+|0010\rangle+|0001\rangle)/2$, it is found numerically that the maximal violation of the inequality (\ref{bi1}) is $2.618$ for some appropriate experimental settings.  We note that both GHZ and W states give rise to the same violation of our inequality (\ref{bi1}) and the SASA inequality, since these states are symmetric under all possible permutations of the qubits.  So, in these cases, interchanging Alice and Bob does not affect their maximal violation.  However, this is nontrivial for the four-qubit cluster state \cite{cluster}
\begin{equation}
|\phi\rangle = \frac{1}{2}
(|+\rangle|0\rangle|+\rangle|0\rangle + |+\rangle|0\rangle|-\rangle|1\rangle +
 |-\rangle|1\rangle|-\rangle|0\rangle + |-\rangle|1\rangle|+\rangle|1\rangle),
\label{clusterphi}
\end{equation}
where $|\pm\rangle \equiv (|0\rangle \pm |1\rangle)/\sqrt{2}$.  $|\phi\rangle$ is not equivalent to $|\chi\rangle$ under stochastic local operations and classical communications (SLOCC) \cite{Osterloh}.  By substituting the following correlation functions
\begin{eqnarray}
Q^{\rm cluster}(A_1B_jC_kD_l) & = & 
\langle\phi|\vec{n}^A_1\cdot\vec{\sigma} \otimes \vec{n}^B_j\cdot\vec{\sigma} \otimes\vec{n}^C_k\cdot\vec{\sigma} 
\otimes\vec{n}^D_l\cdot\vec{\sigma}|\phi\rangle, \nonumber \\
Q^{\rm cluster}(B_jC_kD_l) & = & 
\langle\phi|{\bf 1}^A \otimes \vec{n}^B_j\cdot\vec{\sigma} \otimes \vec{n}^C_k\cdot\vec{\sigma} 
\otimes\vec{n}^D_l\cdot\vec{\sigma}|\phi\rangle
\end{eqnarray}
into the left hand side of the inequality (\ref{bi1}), we can find the maximal value ${\cal B}^{\rm cluster}_{(\ref{bi1})} = 2\sqrt{2}$ predicted by quantum mechanics, for the measurement settings $\vec{n}^A_1 = \vec{x}$; $\vec{n}^B_1 = \vec{z}$, $\vec{n}^B_2 = \vec{z}$; $\vec{n}^C_1 = \frac{1}{\sqrt{2}}(\vec{x}+\vec{z})$, $\vec{n}^C_2 = \frac{1}{\sqrt{2}}(\vec{x}-\vec{z})$; and $\vec{n}^D_1 = \vec{x}$, $\vec{n}^D_2=\vec{z}$.  This clearly demonstrates that the inequality (\ref{bi1}) is not optimally violated by the cluster state.  The entangled state $|\chi\rangle$, being a resource for realizing the teleportation protocol ${\cal E}_0$, can thus be discriminated from $|\phi\rangle$, the GHZ and W states using our inequality (\ref{bi1}).

Last but not least, we note that through cyclic permutations of the four observers, $A \rightarrow B \rightarrow C \rightarrow D \rightarrow A$, we can further obtain two other seemingly different inequalities.  Namely,
\begin{equation}
Q(A_1B_1C_1D_1) + Q(A_2B_2D_1) + Q(A_1B_2D_2) - Q(A_2B_1C_1D_2) \leq 2,
\label{bi3}
\end{equation}
and
\begin{equation}
Q(A_1B_1C_1D_1) + Q(A_1B_2C_2) + Q(A_2B_1C_2) - Q(A_2B_2C_1D_1) \leq 2.
\label{bi4}
\end{equation}
However, inequalities (\ref{bi1}) and (\ref{bi3}) are really of the same kind in the sense that they are optimally violated by $|\chi\rangle$, but $|\phi\rangle$ only yields maximal violation $2\sqrt{2}$ for both.  Similarly, inequalities (\ref{bi2}) and (\ref{bi4}) are of the same type, since $|\phi\rangle$ gives optimal violation and $|\chi\rangle$ yields $2\sqrt{2}$ for both.  The inequalities (\ref{bi1}) and (\ref{bi3}) are optimal for the state $|\chi\rangle$, in the same sense that the inequalities (\ref{bi2}) and (\ref{bi4}) are optimal for the cluster state $|\phi\rangle$.

\section{Bell inequalities versus teleportation}\label{4}
Now, we consider the following four-qubit mixed state, which generalizes the two-qubit Werner state, Eq.(\ref{Werner}), studied by Popescu in Ref.\cite{Popescu}.
\begin{equation}
\Xi(\alpha, \beta) = q|\Upsilon^{00}(\alpha, \beta)\rangle\langle\Upsilon^{00}(\alpha, \beta)| + \frac{1 - q}{16}I_{16},
\label{Xi}
\end{equation}
where $0 \leq q \leq 1$ and $I_{16}$ is the sixteen-dimensional identity.  The generalized singlet fraction is given by \cite{YeoII}
\begin{eqnarray}
{\cal G}[\Xi] & \equiv & 
\max\langle\Upsilon^{00}(\theta_{12}, \phi_{12})|\Xi(\alpha, \beta)|\Upsilon^{00}(\theta_{12}, \phi_{12})\rangle \nonumber \\
& = & \max_{\theta_{12}, \phi_{12}}\frac{1 - q}{16} + \frac{q}{4}[\cos(\theta_{12} - \alpha) + \cos(\phi_{12} - \beta)]^2 \nonumber \\
& = & \frac{1 + 15q}{16},
\end{eqnarray}
when $\theta_{12} = \alpha$ and $\phi_{12} = \beta$.  Clearly, ${\cal G}[\Xi] \leq 1/2$ and $\Xi$ does not yield two-qubit teleportation fidelity better than classical protocol when $q \leq q_{\rm crit}({\cal E}_0) = 7/15$.  When $\alpha = \beta = \pi/4$, we have $\xi = q|\chi\rangle\langle\chi| + (1 - q)/16\ I_{16}$ and critical visibility $q_{\rm crit}({\rm Bell}) = 1/2$ \cite{Comment}.  Hence, there exists $q_{\rm crit}({\cal E}_0) < q < q_{\rm crit}({\rm Bell})$ such that $\xi$ is a useful resource for two-qubit teleportation but nevertheless ``local''.

In order to gain more insight, we consider input states $|\Psi_{\rm in}\rangle = \cos\epsilon|00\rangle + \sin\epsilon|11\rangle$ with $0 \leq \epsilon \leq \pi/2$.  The negativity \cite{Vidal} of the teleported (output) state $\rho_{\rm out}$ is given by
\begin{equation}
{\cal N}[\rho_{\rm out}] = \max\{0,\ -\frac{1}{2}(1 - q) + q\sin2\epsilon\},
\end{equation}
which is zero whenever $q < q^{\bf I}_{\rm crit} \equiv 1/(1 + 2\sin2\epsilon)$.  An equally straightforward calculation yields $q^{\bf II}_{\rm crit} \equiv 1/\sqrt{1 + \sin^22\epsilon}$, such that for $q \leq q^{\bf II}_{\rm crit}$, the output states do not violate the CHSH inequality \cite{Horodecki2}.  Clearly, $q^{\bf II}_{\rm crit} > q^{\bf I}_{\rm crit}$ for all $0 < \epsilon \leq \pi/2$.  This is consistent with the fact that entangled states are not necessarily nonlocal.  Furthermore, we have $q^{\bf II}_{\rm crit} > q_{\rm crit}(\rm Bell) > q_{\rm crit}({\cal E}_0)$; namely, for an output state to remain nonlocal demands that $\xi$ be nonlocal and ``nonclassical''.  More specifically, we pick $\epsilon = \pi/12$, then $q^{\bf I}_{\rm crit} = 1/2$ and $q^{\bf II}_{\rm crit} \approx 0.894427$.  This implies there are $1/2 < q < 0.894427$ such that the output state is entangled but local.   It also means that even when we have nonclassical teleportation fidelity, the entanglement of two-qubit states with entanglement smaller than some critical amount may become zero in ${\cal E}_0$.  These states are being teleported  to separable states with average fidelities that are nevertheless not achievable by ``classical'' means.  Entanglement is fragile to teleport and nonlocality is even more so.

\section{Quantum nonlocality of $|\Upsilon^{00}(\theta_{12}, \phi_{12})\rangle$}\label{5}
\begin{figure}
\begin{center}
\epsfig{figure=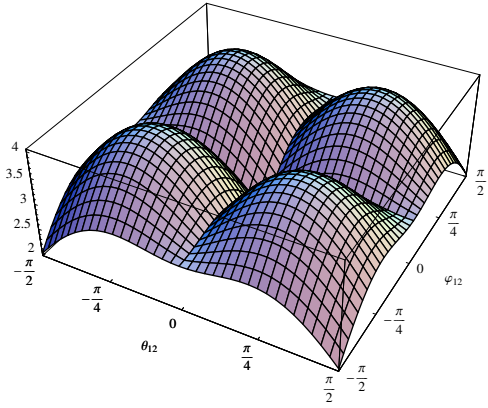,width=0.4\textwidth}
\end{center}
\caption{Numerical results of violation of the inequality (\ref{bi1}) by the states $|\Upsilon^{00}(\theta_{12}, \phi_{12})\rangle$.}
\label{fig}
\end{figure}

In this section, we study the quantum nonlocal property of the state $|\Upsilon^{00}(\theta_{12}, \phi_{12})\rangle$ using our inequality (\ref{bi1}).  The violation of local realism naturally depends on $\theta_{12}$ and $\phi_{12}$.  In Fig. \ref{fig}, we plot the quantum prediction for the Bell quantity constructed from the inequality (\ref{bi1}) for the state $|\Upsilon^{00}(\theta_{12}, \phi_{12})\rangle$ versus $\theta_{12}$ and $\phi_{12}$.  It is shown that the quantum violation varies periodically with $\theta_{12}$ and $\varphi_{12}$.  For a fixed $\theta_{12}$ (or $\phi_{12}$), the quantum violation increases with $\phi_{12}$ (or $\theta_{12}$) from $-\pi/2$ to $-\pi/4$, decreases from $-\pi/4$ to $0$, and then increases again till $\pi/4$ after which it decreases again.

When $\theta_{12}, \phi_{12} \in \{\pm\pi/2\}$, the state $|\Upsilon^{00}(\theta_{12}, \phi_{12})\rangle$ does not violate the inequality (\ref{bi1}).  In each of these cases, $|\Upsilon^{00}(\theta_{12}, \phi_{12})\rangle$ reduces to a tensor product of two Bell states, which though is an entangled state is not a {\em genuine} four-qubit entangled state \cite{Osterloh}.  This means that our Bell inequality is not very strong in detecting this kind of entanglement.

The maximum violation $4$ is obtained when $\theta_{12} = \phi_{12} = \pm\pi/4$, or $\theta_{12} = -\phi_{12} = \pm\pi/4$.  $|\Upsilon^{00}(\pi/4, \pi/4)\rangle = |\chi\rangle$ and $|\Upsilon^{00}(-\pi/4, -\pi/4)\rangle$ is local unitarily equivalent to $|\chi\rangle$.  For the other two cases, we have
\begin{equation}
|\Upsilon^{00}(\pi/4, -\pi/4)\rangle = \frac{1}{2\sqrt{2}}
(|0000\rangle - |0011\rangle + |0101\rangle + |0110\rangle + |1001\rangle - |1010\rangle + |1100\rangle + |1111\rangle),
\end{equation}
and
\begin{equation}
|\Upsilon^{00}(-\pi/4, \pi/4)\rangle = \frac{1}{2\sqrt{2}}
(|0000\rangle + |0011\rangle - |0101\rangle + |0110\rangle + |1001\rangle + |1010\rangle - |1100\rangle + |1111\rangle),
\end{equation}
which are clearly local unitarily equivalent to each other, and are SLOCC equivalent to the state $|\chi\rangle$ \cite{Osterloh}.  They are equally good resources for two-qubit teleportation via ${\cal E}_0$, and our inequality (\ref{bi1}) is efficient at detecting them. 

\section{Conclusions}\label{6}
In conclusion, we have derived a new four-qubit Bell inequality (\ref{bi1}).  Using our inequality, we study the nonlocal quantum properties of several four-qubit states, such as the GHZ and W states, the cluster state $|\phi\rangle$, and the state $|\chi\rangle$ [Eq.(\ref{chi})].  It is shown that, while it is not violated by the four-qubit GHZ state, $|\chi\rangle$ yields optimal violation of the inequality.  We show that our inequality is violated by $|\phi\rangle$, though not optimally.  It can thus be used to detect the state $|\chi\rangle$ experimentally.  In particular, it can be used to discriminate between $|\chi\rangle$ and $|\phi\rangle$.  This has application in ascertaining if a source is emitting the necessary four-qubit entangled states $|\chi\rangle$ for two-qubit teleportation using ${\cal E}_0$.  We consider the violation degree, as measured by the critical visibility, of our inequality by $|\chi\rangle$ and $|\phi\rangle$; and explore the different aspects of four-partite entanglement by considering the usefulness of the state $\xi$ [Eq.(\ref{Xi})] as resource for two-qubit teleportation.  We show that there are four-qubit mixed states that are local but yet are useful resource for two-qubit teleportation, and thus generalize the results obtained in Ref.\cite{Popescu}.  The quantum nonlocality of a general genuine four-qubit entangled state $|\Upsilon^{00}(\theta_{12}, \phi_{12})\rangle$, which includes the state $|\chi\rangle$ as a special case, is also investigated using the inequality (\ref{bi1}).  It is shown that the quantum violation varies periodically with $\theta_{12}$ and $\varphi_{12}$.  We hope that our results would throw more light on the very interesting subject of multipartite entanglement.

\section{Acknowledgement}
This work is supported by NUS academic research Grant No. WBS: R-144-000-123-112.  C.-F. Wu acknowledges financial support from Singapore Millenium Foundation.


\begin{thebibliography}{99}
\bibitem{Schrodinger} E. Schrodinger, Proc. Cambridge Philos. Soc. {\bf 31}, 555 (1935).
\bibitem{Bell} J. S. Bell, Physics (Long Island City, N. Y.) {\bf 1}, 195 (1964).
\bibitem{CHSH} J. F. Clauser, M. A. Horne, A. Shimony, and R. A. Holt, Phys. Rev. Lett. {\bf 23}, 880 (1969).

\bibitem{Nielsen} M. A. Nielsen and I. L. Chuang, {\it Quantum Computation and Quantum Information} (Cambridge University Press, Cambridge, 2000).

\bibitem{Bennett} C. H. Bennett, G. Brassard, C. Crepeau, R. Jozsa, A. Peres, and W. K. Wootters, 
Phys. Rev. Lett. {\bf 70}, 1895 (1993).
\bibitem{Bouwmeester} D. Bouwmeester, J. -W. Pan, K. Mattle, M. Eibl, H. Weinfurter, and A. Zeilinger, Nature {\bf 390}, 575 (1997).

\bibitem{Terhal} B. M. Terhal, Phys. Lett. A {\bf 271}, 319 (2000).

\bibitem{Popescu} S. Popescu, Phys. Rev. Lett. {\bf 72}, 797 (1994).
\bibitem{Werner} R. F. Werner, Phys. Rev. A {\bf 40}, 4277 (1989).
\bibitem{Horodecki1} M. Horodecki, P. Horodecki, and R. Horodecki, Phys. Rev. A {\bf 60}, 1888 (1999).
\bibitem{Horodecki2} R. Horodecki, P. Horodecki, and M. Horodecki, Phys. Lett. A {\bf 200}, 340 (1995).

\bibitem{YeoI} Y. Yeo and W. K. Chua, Phys. Rev. Lett. {\bf 96}, 060502 (2006).
\bibitem{YeoII} Y. Yeo, Phys. Rev. A {\bf 74}, 052305 (2006).

\bibitem{Verstraete} F. Verstraete, J. Dehaene, B. De Moor and H. Verschelde, Phys. Rev. A {\bf 65}, 052112 (2002).
\bibitem{Osterloh} A. Osterloh and J. Siewert, Phys. Rev. A {\bf 72}, 012337 (2005).

\bibitem{Mermin90} N. D. Mermin, Phys. Rev. Lett. {\bf 65}, 1838 (1990).
\bibitem{Ardehali} M. Ardehali, Phys. Rev. A {\bf 46}, 5375 (1992).
\bibitem{BK} A. V. Belinskii and D. N. Klyshko, Phys. Usp. {\bf 36}, 653 (1993).

\bibitem{Greenberger} D. M. Greenberger, M. A. Horne and A. Zeilinger, in {\it Bell's Theorem, Quantum Theory, and Conceptions of the Universe}, edited by M. Kafatos (Kluwer Academic, Dordrecht, 1989), pp. 69-72.

\bibitem{cluster} V. Scarani, A. Ac\'{i}n, E. Schenck and M. Aspelmeyer, Phys. Rev. A {\bf 71}, 042325 (2005).
\bibitem{Briegel} H.-J. Briegel and R. Raussendorf, Phys. Rev. Lett. {\bf 86}, 910 (2001).

\bibitem{Zeilinger} A. Zeilinger, M. A. Horne, and D. M. Greenberger, in
{\it Proceedings of Squeezed States and Quantum Uncertainty}, edited by D. Han, Y. S. Kim, and W. W. Zachary,
NASA Conference Publication No. 3135 (NASA, Washington, DC,  1992), pp. 73-81.

\bibitem{werner} R. F. Werner and M. M. Wolf, Phys. Rev. A {\bf 64}, 032112 (2001).
\bibitem{ZB} M. \.{Z}ukowski and \v{C}. Brukner, Phys. Rev. Lett. {\bf 88}, 210401 (2002).
\bibitem{3qubit} J. L. Chen, C. F. Wu, L. C. Kwek and C. H. Oh, Phys. Rev. Lett. {\bf 93}, 140407 (2004).

\bibitem{Comment} The corresponding critical visibility for the cluster state turns out to be $1/\sqrt{2}$.  This means that, with respect to the inequality (\ref{bi1}), $|\chi\rangle$ is more highly resistant to noise than the cluster state.

\bibitem{Vidal} G. Vidal and R. F. Werner, Phys. Rev. A {\bf 65}, 032314 (2002).
\end{thebibliography}
\end{document}